\documentclass[submission,copyright,creativecommons]{eptcs}
\providecommand{\event}{Physics and Computation 2018} 
\usepackage{breakurl}             
\usepackage{amssymb,amsmath,graphicx,setspace,color,cite,sectsty}
\title{An Atemporal Model of Physical Complexity}
\author{Richard Whyman
\institute{The University of Leeds,\\
Leeds, UK}
\email{mmrajw@leeds.ac.uk}
}
\date{}
\def\titlerunning{An Atemporal Model of Physical Complexity}
\def\authorrunning{Richard Whyman}
\newcounter{ThmNum}

\newtheorem{thrm}{Theorem}[section]
\newtheorem{defn}{Definition}[section]
\newtheorem{exmpl}{Example}[section]
\newtheorem{rmrk}{Remark}[section]
\begin{document}
\maketitle
\begin{abstract}
We present the finite first-order theory (FFOT) machine, which provides an atemporal description of computation. We then develop a concept of complexity for the FFOT machine, and prove that the class of problems decidable by a FFOT machine with polynomial resources is $NP \cap \text{co-}NP$.
\end{abstract}

\noindent
In the 1960's Cobham and Edmonds \cite{cobham1969intrinsic,edmonds1965paths} asserted that a computational problem is feasibly computable if and only if it can be decided in polynomial time on a Turing machine (and thereby lies in $P$). Though not within the originally intended scope, it has been suggested \cite{vergis1986complexity} that Cobham and Edmonds assertion should also apply to what is feasibly computable by \textsl{any} physical system \footnote{By physical system, we mean anything whose physical properties are known that we may realistically put together and use. Computing with a physical system should then involve being able to input data into the system (by adjusting the locations or properties of the objects within) so that we may reliably observe an output from the system that provides a solution to some problem of ours. Examples include a table of ball bearings and grooves, a screen diffracting a ray of light, a slide rule, and indeed a normal digital computer.}. 
However, this idea has since been challenged by results from quantum computation \cite{nielsen2010quantum}, such as Shor's factorisation algorithm, which suggest that the class of problems decidable by a quantum computer in polynomial time ($BQP$) may include problems that do not lie in $P$. These results lead naturally to the questions of what it is about quantum systems that makes them capable of feasibly deciding problems outside of $P$, and whether there exist other physical systems with such capabilities. 

In \cite{Blakey2010model} Blakey described a collection of classical physical devices capable of factorising integers in polynomially bounded space and time. However, Blakey argued that, unlike quantum factorisers, his factorisation systems are not feasibly realisable, as the \textsl{precision} required to implement such a system has to grow exponentially with the size of the input. Blakey then went on to assert that in general the resource usage of physical computational devices should be measured in more than just time and space alone. For example the energy or precision required by a computation should also be considered. 

In \cite{baumeler2018computational} Baumeler and Wolf looked into the computational power of polynomially bounded circuits acting within closed timelike curves of polynomial length. They asserted that a computation may occur on such a circuit if it is logically consistent and unique, demonstrating that with these assumptions the computational power of these non-causal circuits is equal to $UP \cap \text{co-}UP$ \footnote{A problem is in $UP$ if it is decidable by a non-deterministic Turing machine in polynomial time, and there exists at most one accepting path for each input.}. Notably, $BQP$ problems such as the factorisation problem also lie in $UP \cap \text{co-}UP$, suggesting that there may be a non-causal aspect to the quantum speed-up. 

Baumeler and Wolf's innovative non-causal circuit model did not have the goal of describing the feasible computational aspects of a general physical system. 
However, in an attempt to do just that, in this paper we develop the concept of a \textbf{finite first-order theory (FFOT) machine}, and describe what it means to  compute efficiently with such a device. In \cite{whyman2018physical} we introduced the concept of a \textbf{theory machine}, which is inspired by Horsman et al.'s reasoning on physical computation \cite{WhenDoesAPhysicalSystemCompute} and Gurevich's sequential abstract-state machines \cite{Gurevich2000}. Rather than describing each computation as a discrete ordered sequence of structures, in a theory machine the whole computation is described via a single consistent structure. Hence any temporal evolution of the machine is described within this structure.  
The inclusion of the evolution within the structure allows a theory machine to compute in a consistent non-causal or atemporal manner. 

In \cite{whyman2018physical} we demonstrated how various super-Turing systems\footnote{By super-Turing we mean a system which is capable of deciding problems that are not decidable by a Turing machine.} are examples of theory machines. Such systems include Blum-Shub-Smale machines \cite{blum1989theory}, which perfectly perform algebraic operations in single time steps, and infinite time Turing machines \cite{hamkins2000infinite}, whose computations take an infinite amount of time steps. Whether such powerful systems should be viewed as physical systems is of course highly questionable, which is why in \cite{whyman2018physical} we also introduced the FFOT machine. FFOT machines are theory machines that are restricted to finitely describing computational systems using only first-order logic. In \cite{whyman2018physical} we proved that a word function is computable by a FFOT machine iff it is computable by a Turing machine. Meaning that the computational aspects of a super-Turing system cannot be described by a FFOT machine.

In \textsl{When does a physical system compute?} \cite{WhenDoesAPhysicalSystemCompute} Horsman, Stepney, Wagner, and Kendon put forward a minimal collection of requirements that a physical system must satisfy in order for it to be capable of computation. Horsman et al. asserted that in order for a person to be able to compute with a physical system they must be able to abstractly represent the necessary workings of the system, whilst possessing a sufficiently correct theory of how the system behaves. We assert that this representation and theory can be expressed in terms of first-order logical sentences. A FFOT machine is then given by a triple $\mathbf{M} = (\mathbf{T},\mathbf{I},\mathbf{O})$ where $\mathbf{T}$ is a set of first-order sentences, and $\mathbf{I}$ and $\mathbf{O}$ are sets of sets of first-order sentences. The theory of the system is given by $\mathbf{T}$, which describes the necessary aspects of a system that we wish to compute with. $\mathbf{T}$ is also finite in order to conform with Horsman et al.'s assertion that the theory must be knowable to the user. The set of admissible inputs into the system is given by $\mathbf{I}$, and the set of measurable outputs from the system is given by $\mathbf{O}$.

The key idea behind a FFOT machine is that for any $\Phi \in \mathbf{I}$ we can obtain a structure $\mathfrak{P}$ which satisfies $\mathbf{T} \cup \Phi$. In $\mathfrak{P}$ there is at most one true output $\Theta \in \mathbf{O}$, we then take $\Theta$ to be the outcome of the computation by $\mathbf{M}$ on input $\Phi$. This structure $\mathfrak{P}$ does not need to contain a clear notion of time, nor does $\Theta$ need to follow from $\mathbf{I}$ via a clear sequence of steps. Hence the typical notion of a sequential causal computation does not necessarily occur within a FFOT machine. However,
we shall insist that the only way $\Theta$ can be the output of $\mathbf{M}$ on input $\Phi$ is if $\Theta$ is true in \textsl{every} model $\mathbf{T} \cup \Phi$, which ensures that the computation can not just happen in one uncomputable step. Instead, as we shall see, the computation must still have a non-trivial amount of structure to it in order to produce an output. 

The nature of a FFOT machine computation is intended to mimic what happens when we use a physical system to carry out a decision process. For example, suppose we wish to compute with some kinematic system of billiard balls, to do this we can use the axioms of Newtonian mechanics as our theory $\mathbf{T}$ to predict the motions of the system (Newtonian mechanics may not be a perfect description of reality, but in many cases it is more than good enough). Each input $\Phi \in \mathbf{I}$ could be a non-contradictory description of the positions and velocities of the balls at some initial time $t_0$. Whereas each output $\Theta \in \mathbf{O}$ could be a position measurement at some final time $t_1$. 
As this is a real physical situation we should always be able to create a kinematic scenario from $t_0$ to $t_1$ in which $\mathbf{T} \cup \Phi$ is satisfied.
Though due to imprecision and the inexactness of the theory $\mathbf{T}$ there are likely to be many scenarios that satisfy $\mathbf{T} \cup \Phi$, however if we know that in each of them only the output $\Theta$ is true, then the exact scenario created does not matter, all that matters is which element of $\mathbf{O}$ is true given an input of $\Phi$.

\section{FFOT machine computation}
Given all of this, below  we give the definition of a FFOT machine in a (first-order) vocabulary $\mathbf{V}$. Where as in \cite{Gurevich2000} a vocabulary consists of a finite set of relations, functions and constant symbols. A $\mathbf{V}$-sentence is then a first-order formula constructed from the elements of $\mathbf{V}$ without any free variables. 
\begin{defn}
\label{TMD}
\normalfont
Let $\mathbf{V}$ be a vocabulary, a \textbf{finite first-order theory (FFOT) machine} in the vocabulary of $\mathbf{V}$ is a triple 
$\mathbf{M} = (\mathbf{T},\mathbf{I},\mathbf{O})$ where $\mathbf{T}$, $\mathbf{I}$ and $\mathbf{O}$ are sets of $\mathbf{V}$-sentences such that:
\begin{itemize}
\item $\mathbf{T}$ is a finite set of $\mathbf{V}$-sentences,
\item $\mathbf{I}$ and $\mathbf{O}$ are sets of sets of $\mathbf{V}$-sentences,
\item For every $\Phi \in \mathbf{I}$ the set $\mathbf{T} \cup \Phi$ is satisfiable,
\item For every $\Phi \in \mathbf{I}$ and $\Theta,\Psi \in \mathbf{O}$ if $\Theta \not= \Psi$ then the set $\mathbf{T} \cup \Phi \cup \Theta \cup \Psi$ is not satisfiable.
\end{itemize}
\vspace{1mm}
We call $\mathbf{T}$ the \textbf{theory} of $\mathbf{M}$, call $\mathbf{I}$ the set of \textbf{inputs} of $\mathbf{M}$, and call $\mathbf{O}$ the set of \textbf{outputs} from $\mathbf{M}$.
We say that the FFOT machine $\mathbf{M}$ \textbf{computes} $\Theta$ from $\Phi$ if $\mathbf{T} \cup \Phi \models \Theta.$
We denote this by: 
$$\mathbf{M}(\Phi) = \Theta.$$
Let $\Theta,\Psi \in \mathbf{O}$ where $\Theta \not= \Psi$, if there exists a model of $\mathbf{T} \cup \Phi$ where $\Theta$ is true and another model of $\mathbf{T} \cup \Phi$ where $\Psi$ is true, then $\mathbf{M}$ cannot compute anything on input $\Phi$ and $\mathbf{M}(\Phi)$ is undefined.
\end{defn}
We believe that the computational aspects of any physical system can be described by FFOT machine, however there may well exist FFOT machines that do not have any physical realisation. 
\begin{exmpl}
\normalfont
Let $\mathbf{V} = \{R,f,c\}$ where $R$ is a unary relation, $f$ a unary function, and $c$ a constant. A simple example of a FFOT machine is $\mathbf{M} = (\mathbf{T},\mathbf{I},\mathbf{O})$ where:
\begin{itemize}
\item $\mathbf{T} = \{\forall x (R(x) \leftrightarrow R(f(x)))\}$,
\item $\mathbf{I} = \{\{R(c)\},\{\neg R(c)\}\}$,
\item $\mathbf{O} = \{R(f(c)),\neg R(f(f(c)))\}$.
\end{itemize}
We then have $\mathbf{M}(\{R(c)\}) = \{R(f(c))\}$ as in any model of $\mathbf{T}$, if $R(c)$ is true then $R(f(c))$ must also be true, so $\mathbf{T} \cup \{R(c)\} \models \{R(f(c))\}$. Whereas $\mathbf{M}(\{\neg R(c)\}) = \{\neg R(f(f(c)))\}$ as given $\neg R(c)$ by $\mathbf{T}$ we then have $\neg R(f(c))$ is true and so $\neg R(f(f(c)))$ is true, hence $\mathbf{T} \cup \{\neg R(c)\} \models \{\neg R(f(f(c)))\}$.
\end{exmpl}
\noindent
As many examples of computational systems write their inputs and outputs as words, we naturally require a standard manner in which to write words as first-order sentences. We can do this by assigning the values of a well-behaved sequence of ground terms \cite{epstein2011classical} to the symbols in the word. 

\begin{defn}
\normalfont
We call a sequence of distinct ground terms $\{\chi_i\}_{i \in \mathbb{N}}$ a \textbf{simple sequence} if every sequent is of the form $\chi_i = \gamma(\sigma^i(\delta))$ where $\delta$ is a ground term, and $\gamma(y)$ and $\sigma(y)$ are terms with a single free variable $y$.

Let $X = (\chi_i)_{i \in \mathbb{N}}$ be a simple sequence. For a set of constants $\Sigma$ with $\mathbf{b} \not\in \Sigma$, the \textbf{$X$-word set} corresponding to $w = w_0w_1 \cdots w_n \in \Sigma^*$ is:
$$\Phi_{X}^w = \bigcup_{i=0}^n \{\chi_i = w_i\} \cup \{\chi_{n+1} = \mathbf{b}\}.$$
We denote the set of $X$-word sets from an alphabet $\Sigma$ by $\hat{\Sigma}^*_{X} = \{\Phi_{X}^w \text{ }|\text{ } w \in \Sigma^*\}.$
\end{defn}
So a finite word set $\Phi_{X}^w$ maps each term $\chi_i$ to the $i$th symbol in $w$, the symbol $\mathbf{b}$ is then intended to represent the ``blank" symbol, 
hence $\chi_{n+1} = \mathbf{b}$ implies that this is the end of the word. This is necessary as without the blank assignment it would be the case that for any prefix word $v$ of $w$ we would have $\Phi_X^{w} \models \Phi_{X}^v$, which would clearly interfere with our notion of computation.
Note that if $\chi_i = \gamma(\sigma^i(\delta))$ then by adding the sentence $\forall y ((\gamma(y) = \mathbf{b}) \rightarrow (\gamma(\sigma(y)) = \mathbf{b}))$ to the theory of a machine with inputs from $\hat{\Sigma}^*_{X}$ we can ensure that $\chi_j = \mathbf{b}$ for each $j > n$.

\begin{rmrk}
\normalfont
For simplicity, $X$-words use the equality symbol ``$=$" in their construction. 
In any FFOT machine with vocabulary $\mathbf{V}$ which takes such inputs we will ensure that $=\, \in \mathbf{V}$ satisfies the usual equality axioms $EQ^=_\mathbf{V}$ (Definition \ref{EAD} in the appendix) of being an equivalence relation which preserves the functions and relations of $\mathbf{V}$. As a FFOT machine's theory is finite, its vocabulary can assumed to be as well, meaning that the equality axioms for $\mathbf{V}$ can always form a finite part of the machine's theory.
\end{rmrk}

\begin{exmpl}
\label{TMFFOTME}
\normalfont
Let $N$ be a Turing machine which decides the problem $A \subseteq \Sigma^*$. For simplicity, we shall take $N$'s tape to be infinite in only the rightwards direction, with a symbol $\mathbf{L}$ marking its leftmost tape cell. Let $N$ use the alphabet $\Lambda \supseteq \Sigma \cup \{\textbf{L},\mathbf{b}\}$, where $\mathbf{b}$ indicates a blank tape cell. Let $N$ use the set of internal states $\Pi$ with initial state $s_0$, accepting state $s_a$, and rejecting state $s_r$. 
Let $N$ follow the set of rules $\mathbf{R}$, each of the form: 
$$(t,b;u,c,p) \in (\Pi \setminus \{s_a,s_r\}) \times \Lambda \times \Pi \times \Lambda \times \{LEFT,PAUSE,RIGHT\},$$
which is read as ``if the machine is in internal state $t$ reading $b$ then go to state $u$, replace the symbol being pointed to with $c$, and move $p$." To avoid the situation where no rule may be applied prior to halting we let $\mathbf{R}$ contain a rule beginning with $(t,b)$ for every $t \in \Pi \setminus \{s_1\}$ and $b \in \Lambda$.
We can then describe $N$ by the FFOT machine: 
$$\mathbf{T}\mathbf{M}_{N} = (\mathbf{T}\mathbf{M}\mathbf{T}_{N},\hat{\Sigma}^*_{\mathbf{X}},\{\{I(h) = s_a\},\{I(h) = s_r\}\}),$$
in the vocabulary of $\mathbf{V}^{N} = \{S,0,C,I,H,h,\textbf{L}\} \cup \Lambda \cup \Pi$. 

Where $\{S,0\}$ are the usual symbols of Peano arithmetic with $S$ as the successor function, $C$ is a binary function, $I,H$ are unary functions, and the rest of the symbols are constants. $C(x,y)$ maps to the constants of $\Sigma$ to describe the contents of the $y$th tape cell at time $x$, whereas $I(x)$ gives the internal state at time $x$ by mapping to the constants of $\Pi$, and $H(x)$ maps to the head position at time $x$. The halting time is represented by the constant symbol $h$, its value depends on the input.
We encode the input words from $\Sigma^*$ via the simple sequence $X = \{C(0,(S^{i+1}(0))\}_{i \in \mathbb{N}}$. The theory of $\mathbf{T}\mathbf{M}_{N}$ is:
$$\mathbf{T}\mathbf{M}\mathbf{T}_N = \left\{\begin{array}{l} (H(0) = 1) \land (C(0,0) = \textbf{L}) \land (I(0) = s_0), \\
\forall y (C(0,y) = \mathbf{b}) \rightarrow (C(0,S(y)) = \mathbf{b}), \\
\forall y (\neg(H(x) = y) \rightarrow (C(S(x),y) = C(x,y)))
\end{array}\right\} \cup EQ^=_{\mathbf{V}^{N}} \cup PSA \cup \mathbf{\hat{R}} \cup HT_{(s_a,s_r)}.$$
Where $PSA$ denotes the set of first-order Peano axioms (Definition \ref{FOPAD} in the appendix). This together with the equality axioms $EQ^=_{\mathbf{V}^{N}}$ ensures that any model $\mathfrak{A}$ of $\mathbf{T}\mathbf{M}\mathbf{T}_N \cup \Phi_X^w$ is an expansion of either the usual structure of the natural numbers $\langle \mathbb{N};=,S,0\rangle$ or a structure with an initial segment that is isomorphic to $\langle \mathbb{N};=,S,0\rangle$ \cite{kaye1991}. The first two sentences in $\mathbf{T}\mathbf{M}\mathbf{T}_N$ together with $\Phi_X^w$ give $\mathfrak{A}$ the initial configuration of $N$ on input $w$. The evolution of the machine in $\mathfrak{A}$ is then given by the third sentence and $\hat{\mathbf{R}}$ where:
$$\mathbf{\hat{R}} = \bigcup_{(t,b;u,c,p) \in \mathbf{R}}\left\{\forall x \big(\mu_{(t,b)}(x,H(x)) \rightarrow \big(\mu_{(u,c)}(S(x),H(x)) \land \pi_{(p)}(H(x),H(S(x)))\big)\big)\right\}.$$
Each sentence of $\hat{\mathbf{R}}$ implements a rule of $\mathbf{R}$ via the term:
$$\mu_{(s,a)}(z_1,z_2) \equiv ((I(z_1) = s) \land (C(z_1,z_2) = a)),$$
which indicates that at time $z_1$ the internal state is $s$, and the cell $z_2$ contains an $a$, as well as the term:
$$\pi_{(p)}(z_1,z_2) \equiv \left\{\begin{array}{cl} 
z_2 = S(z_1) & \quad \text{if } p = \text{RIGHT}, \\ 
z_2 = z_1 & \quad \text{if } p = \text{PAUSE}, \\ 
S(z_2) = z_1 & \quad \text{if } p = \text{LEFT},
\end{array}\right.$$
which indicates how tape cell $z_1$ relates to tape cell $z_2$. The remaining sentences of $\mathbf{T}\mathbf{M}\mathbf{T}_N$ are:
$$HT_{(s_a,s_r)} = \left\{\begin{array}{ll}
\forall x ((I(S(x)) = s_a) \land \neg(I(x) = s_a)) \rightarrow (h = S(x)), & \forall x ((I(x) = s_a) \rightarrow (I(S(x)) = s_a)), \\
\forall x ((I(S(x)) = s_r) \land \neg(I(x) = s_r)) \rightarrow (h = S(x)), & \forall x ((I(x) = s_r) \rightarrow (I(S(x)) = s_r))
\end{array}\right\} $$
which ensure that $h$ is the first time step of $\mathfrak{A}$ at which the machine is at either the accepting state or the rejecting state, and afterwards $\mathfrak{A}$ remains in that state. 

By assumption, $N$ eventually accepts or rejects any input $w$, which means that either $s_a$ or $s_r$ must be reached at some finite time step, hence $h$ is necessarily located in the initial segment of $\mathfrak{A}$. Therefore the value of $I(h)$ is entirely determined prior to $h$, and anything that occurs after $h$ or at non-standard time steps cannot affect this output without leading to an inconsistent model. Consequently we have that $\mathbf{T}\mathbf{M}_{N}(\Phi_X^w) = \{I(h) = s_a\}$ iff $N$ accepts $w \in \Sigma^*$.
\end{exmpl}
\begin{exmpl}
\label{DSFE}
\normalfont
As noted in the introduction, in \cite{Blakey2010model} Blakey described a classical physical device capable of factorising integers in polynomially bounded space and time. His device consists of a screen with a pair of slits of distance 1 apart with a light source placed halfway between the two slits. It also includes a detector that runs perpendicular to the screen from one of the slits which is able detect sufficiently strong instances of radiation at integer distances from the screen. To factorise the integer $n \in \mathbb{N}$ one makes the light source emit radiation of wavelength $\frac{1}{2\sqrt{n}}$, the two slits then diffract the light and cause interference pattern on the detector. Blakey showed that if maximal constructive interference is detected at a distance $h$ from the screen then  $\sqrt{n}(\sqrt{h^2+1}+h)$ must be a factor of $n$.

Blakey's double slit factorisation system can be described by an electromagnetic wave function $\mathbf{E}: \mathbb{R}^4 \rightarrow \mathbb{R}^3$ whose propagation depends on the electromagnetic wave equations and the constraints detailed above. Such a description can then be implemented by a FFOT machine $\mathbf{M}$ which satisfies the first-order axioms of a dense ordered field \cite{vakil2011real} (Definition \ref{FORAD} in the appendix). These axioms are modelled by the usual structure of the reals $\langle \mathbb{R};=,<,\leqslant,+,\times,0,1\rangle$, so $\mathbf{E}$ can be described via quaternary functions $E_1,E_2,E_3$. It is then possible to define the partial derivatives of these functions. Typically the partial derivative of $E_i$ in the 1st dimension is defined to be: 
$$\partial_1 E_i(x,y,z,t) = \lim_{\delta \rightarrow 0}\frac{|(E_i(x + \delta,y,z,t) - E_i(x,y,z,t)|}{\delta}.$$
Hence we can define this in the vocabulary of the machine as a quaternary function $\partial_1 E_i$ which satisfies:
$$\forall x \forall y \forall z \forall t \forall \epsilon \exists \delta (((0 < \epsilon) \land (0 < \delta)) \rightarrow ((|((E_i(x + \delta,y,z,t) - E_i(x,y,z,t)) - (\partial_1 E_i(x,y,z,t) \times \delta))| \leqslant \epsilon)).$$
Therefore the electromagnetic wave equations can then be implemented in $\mathbf{M}$ by explicitly writing them out in the vocabulary of $\mathbf{M}$ in the theory of $\mathbf{M}$.
The screen and slits can be implemented as boundary conditions whereas the location of the light source $l$ may be specified by an input of the form $\Phi_Y^{w}$ for $w \in \{0,1\}^*$ and $Y = \{B(D^k(n))\}_{k \in \mathbb{N}}$. Where $B$ and $D$ are unary functions such that $D(y) = \frac{y}{2}$, and:
$$B(y) = \left\{\begin{array}{ll} 
0 & \text{if } y \in \bigcup_{m=1}^\infty[2m,2m + 1),\\ 
1 & \text{if } y \in \bigcup_{m=1}^\infty[2m + 1,2m),\\ 
\mathbf{b} & \text{if } y \in [0,1).\end{array}\right.$$

It is then the case that $B(D^k(y))$ gives the $k$th binary digit of $n$ (reading from right to left). We can then define $l$ within the theory as satisfying $(l \times l \times (1 + 1 + 1 + 1) \times n) = 1$. The output can be extracted via a similar mechanism.

There is a potential problem though, the dense ordered field axioms do not include the second-order least upper-bound axiom, which means that they are also satisfied by the usual structure of the rationals $\langle \mathbb{Q};=,<,\leqslant,+,\times,0,1\rangle$, as well as various non-standard models. However, as in Example \ref{TMFFOTME} these other possible models will still give the correct output. Blakey's device was designed to output correctly even if there is a degree of error, so a rational model is not a problem, whereas any non-standard elements should be unable to interact with the rest of the model.
\end{exmpl}
\noindent
Quantum computers and fluid-based computers may also be described by FFOT machines, details of how this can be done may be found in \cite{whyman2018physical}.

\begin{defn}
\label{PDD}
\normalfont
Let $A \subseteq \Sigma^*$ be a word problem. We say that a FFOT machine $\mathbf{M} = (\mathbf{T},\mathbf{I},\mathbf{O})$ in the vocabulary of $\mathbf{V}$ is \textbf{able to compute} $A$ if there exists a simple sequence $X$ such that $\hat{\Sigma}^*_{X} \subseteq \mathbf{I}$, and for two distinct finite output sets $\Theta,\Psi \in \mathbf{O}$ we have that for every $w \in \Sigma^*$:
$$(w \in A \iff \mathbf{M}(\Phi_{X}^w) = \Theta) \text{ }\text{ and }\text{ } (w \not\in A \iff \mathbf{M}(\Phi_{X}^w) = \Psi).$$
\end{defn}
So a FFOT machine is able to compute a word function if there exists a
way in which we can configure each input word into the machine, such that the output of the function can clearly determined from the machine. Note that a problem can only be computed by a FFOT machine if every possible input word can be encoded into the machine, as we should not be able to just ignore troublesome inputs.
\begin{thrm}
\label{FTOTM}
A word problem $A \subseteq \Sigma^*$ is computable by a Turing machine if and only if there exists a finite first-order theory machine which is able to compute $A$.
\end{thrm}
\textsl{Proof:}
$(\Rightarrow)$ By Example \ref{TMFFOTME} if $A \subseteq \Sigma^*$ is computed by a Turing machine $N$ then the FFOT machine $\mathbf{T}\mathbf{M}_{N}$ is able to compute $A$ via the simple sequence $X$ and outputs $\{I(h) = s_a\}$ and $\{I(h) = s_r\}$.

$(\Leftarrow)$ This follows from the fact that for any FFOT machine $\mathbf{M} = (\mathbf{T},\mathbf{I},\mathbf{O})$ and any input $w$ encoded as $\Phi_X^w$ we must have $\mathbf{T} \cup \Phi_X^w \models \Theta$ for some $\Theta \in \mathbf{O}$. As first order logic is complete there must therefore exist a finite proof of each element of $\Theta$ from $\mathbf{T} \cup \Phi_X^w$, which can be found by enumerating all proofs from $\mathbf{T} \cup \Phi_X^w$ and halting when the entirety of an element of $\mathbf{O}$ is found. (A full proof of this direction can be found in \cite{whyman2018physical}.) \hfill $\Box$
\vspace{2mm}

\noindent
The (generally accepted) Church-Turing thesis \cite{cooper2003computability,deutsch1985quantum} states that \textsl{``Every effectively calculable function is computable by a Turing machine"}. The Church-Turing thesis was originally only meant to assert that anything a \textsl{person} is able to calculate is computable by a Turing machine, however it has been suggested \cite{deutsch1985quantum} that it also applies to what we may effectively calculate via a physical system. Consequently if the Church-Turing thesis does apply to physical computation then by the above result the computational capabilities of any usable physical system must be describable by a FFOT machine. 
\section{FFOT machine complexity} 
Though a Turing machine is typically defined as being unbounded in time and space, a halting computation on a Turing machine is usually understood to be finite in time and space. Hence we may describe a Turing machine computation in time $t$ and space $s$ via a structure with a domain of size Max$(t,s)$. Similarly we may view a kinematic system as a continuously infinite structure, but if when implementing it we require only bounded precision (as in Example \ref{DSFE}), along with bounded space and time, then a computation on it may be described by a finite approximating structure. For example a computation of precision $\epsilon$, taking time $t$ and within a space of diameter $r$, may be described via a structure of size Max$(t,\frac{r}{\epsilon})$. 
We therefore argue that if a FFOT machine on input $\Phi$ is satisfied by a finite structure of size $n$, then the amount of computational resources required to carry out a computation on input $\Phi$ is of order at most $n$.
\begin{defn}
\label{QPDD}
\normalfont
Let $A \subseteq \Sigma^*$ be a word problem, and $q: \mathbb{N} \rightarrow \mathbb{N}$ be a strictly increasing function.
We say that a FFOT machine $\mathbf{M} = (\mathbf{T},\mathbf{I},\mathbf{O})$ is able to \textbf{compute $A$ with $q$ resources} if $\mathbf{M}$ is able to compute $A$ via some simple sequence $X$ and $\hat{\Sigma}^*_{X} \subseteq \mathbf{I}$, such that for every $w \in \Sigma^*$ there exists a structure $\mathfrak{A}$ where:
$$\mathbf{T} \cup \Phi_{X}^w \models \mathfrak{A}, \text{ }\text{ and }\text{ } |\textbf{dom}(\mathfrak{A})| \leqslant q(|w|),$$
where $\textbf{dom}(\mathfrak{A})$ denotes the domain of $\mathfrak{A}$.
\end{defn}
So if a physical system can be described by a FFOT machine which is able to compute a problem $A$ with $q$ resources then we believe that such a system requires at most order $q$ resources to decide A.

\begin{exmpl}
\label{TMFMFFOTME}
\normalfont
Despite describing a Turing machine, our FFOT machine in Example \ref{TMFFOTME} cannot compute any problem with a finite amount of resources. This is because
every structure which satisfies the machine's theory is an expansion of $\mathbb{N}$ and therefore infinite.

However it is possible to describe a Turing machine with a FFOT machine that has bounded models of arbitrary finite size, we just need to replace $PSA$ in the theory of $\mathbf{T}\mathbf{M}_{N}$ with $PSA_f$ (Definition \ref{FFOPAD} in the appendix). $PSA_f$ defines a number space similar to $\mathbb{N}$ that has a specified greatest number $e$, with $S(e) = e$. Models of $PSA_f$ include structures with domain $\{0,1,\hdots,n-1,n\}$ for any $n \in \mathbb{N}$. 
We can then describe a Turing machine $N$ as in Example \ref{TMFFOTME} by the FFOT machine: 
$$\mathbf{T}\mathbf{M}'_{N} = (\mathbf{T}\mathbf{M}\mathbf{T}'_{N},\hat{\Sigma}^*_{\mathbf{X}},\{\{I(h) = s_a\},\{I(h) = s_r\}\}),$$ in the vocabulary of $\mathbf{V}^{N} \cup \{e\}$, where $\mathbf{T}\mathbf{M}\mathbf{T}_N' = (\mathbf{T}\mathbf{M}\mathbf{T}_N \setminus PSA) \cup PSA_f$,
and $\mathbf{V}^{N}$ and $\mathbf{T}\mathbf{M}\mathbf{T}_N$ are as they are in Example \ref{TMFFOTME}.
If a computation of $N$ on input $w$ takes $n$ time steps before halting then any model $\mathfrak{B}$ of $\mathbf{T}\mathbf{M}\mathbf{T}'_{N} \cup \Phi^w_X$ must have at least $n$ elements. As by the rules in $\mathbf{\hat{R}}$ the values of either $C(x,y)$, $I(x)$ or $H(x)$ must change moving from time $x$ to time $S(x)$ (if they did not change then $N$ would be stuck in a never halting loop), leading to a contradiction if $S(x) = x$.

Conversely for large enough $n$ we may have $|\textbf{dom}(\mathfrak{B})| = n + 1$ with $\textbf{dom}(\mathfrak{B}) = \{0,1,\hdots,n\}$ and $n = h = e$. As after time $n$ the state of $N$ is either $s_a$ or $s_r$, in which case no rule of $\mathbf{\hat{R}}$ may be applied and there is no need for the values of any of $C(x,y)$, $I(x)$ or $H(x)$ to differ from $C(S(x),y)$, $I(S(x))$ or $H(S(x))$. 

Consequently if a problem $A \subseteq \Sigma^*$ is computable by a polynomial time Turing machine with time function $p: \mathbb{N} \rightarrow \mathbb{N}$, then $A$ is computable by a FFOT machine with $p$ resources.
\end{exmpl}
Unlike exponential growth, a polynomial resource growth is relatively manageable.
We therefore believe that a FFOT machine can feasibly decide a problem if and only it is able to decide the problem with polynomially resources. This fits with the usual notions of what is feasibly computable with other well known models of computation.
\begin{exmpl}
\normalfont
As in Example \ref{TMFMFFOTME} we can convert the FFOT machine describing Blakey's factorisation system in Example \ref{DSFE} into a FFOT machine with finite models. This is done by modifying the dense ordered field axioms to make them have finite models which serve as approximations to $\mathbb{R}$ and $\mathbb{Q}$ (These axioms, $DOF_f$, are given by Definition \ref{FFORAD} in the appendix). 

We can then describe Blakey's factorisation system in an otherwise identical manner to before. Partial derivatives can be defined as approximations to there true value using the same definition given in Example \ref{DSFE}. Since Blakey's factorisation system was designed to output even with a degree of error, the outputs will also be the same, provided that each model of the system is sufficiently precise. To ensure that we have enough precision we can define within the theory the error of the model $\frac{1}{r}$ (Detailed in Definition \ref{FFORAD} in the appendix) to be sufficiently small in relation to the input. For example we may have $(\frac{1}{r} \times n \times n \times n) \leqslant 1$, ensuring that $\frac{1}{r} \leqslant \frac{1}{n^3}$.

Clearly inputting $n$ should, in general, give a different output to inputting $n+1$. Hence there must be a clear separation between $\frac{1}{2\sqrt{n}}$ and $\frac{1}{2\sqrt{n + 1}}$, which means that in order to implement the device the error $\frac{1}{r}$ must be less than $|\frac{1}{2\sqrt{n + 1}} - \frac{1}{2\sqrt{n}}|$. This error shrinks at an inverse polynomial rate with respect to $n$, and at an inverse exponential rate with respect to the length of $n$'s binary expansion. The axioms of $DOF_f$ imply that, between its greatest and least element, the structure is closed under addition, meaning that there are at least $r$ elements between $0$ and $1$ alone. Therefore, as $r$ grows exponentially with the size of the input, so must the minimal model size.

We therefore conclude that such a FFOT machine requires at least exponential resources to compute the factorisation problem, agreeing with Blakey's \cite{Blakey2010model} idea that precision should be viewed as a resource.
\end{exmpl}
\begin{rmrk}
\normalfont
The polynomial time non-causal circuits of Baumeler and Wolf \cite{baumeler2018computational}, may also be described by polynomial resource-bounded FFOT machines. However they are only able to decide problems in $UP \cap \text{co-}UP$, as unlike FFOT machines, each circuit must have a unique satisfying model. Choosing to limit FFOT machines in such a way would give us $UP \cap \text{co-}UP$ in the above result. However doing so would mean that a FFOT machine would have to provide the definitive description of the physical system it is describing, something that may well be impossible to verify.
\end{rmrk}
\begin{thrm}
A problem is computable by a FFOT machine with polynomial resources if and only if it is in $NP \cap \text{co-}NP$.
\end{thrm}
\textsl{Proof:}
$(\Rightarrow)$ Let $p$ be a polynomial function and $\mathbf{M} = (\mathbf{T},\mathbf{I},\mathbf{O})$ be a FFOT machine in the vocabulary $\mathbf{V}$ which computes $A \subseteq \Sigma^*$ with $p$ resources.
So by assumption, for some simple sequence $\mathbf{X}$ and $\Theta,\Psi \in \mathbf{O}$, we have $\Sigma^*_\mathbf{X} \subseteq \mathbf{I}$ and for each $w \in \Sigma^*$ there is a finite $\mathbf{V}$-structure $\mathfrak{A}$ satisfying $\mathbf{T} \cup \Phi^w_\mathbf{X}$ with $|\textbf{dom}(\mathfrak{A})| \leqslant p(|w|)$. Also if $w \in A$ then $\mathfrak{A} \models \Theta$ and if $w \not\in A$ then $\mathfrak{A} \models \Psi$. We can non-deterministically obtain such a structure as follows.

Let $\mathbf{V}$ contain $m$ relations, $k$ functions, and $r$ constant symbols, also let each relation and function have an arity at most $l$. We can encode each element of $\textbf{dom}(\mathfrak{A})$ as a word in $\{0,1\}^{p(|w|)}$. Each relation can then be encoded as a string of length $O(p(|w|)^l)$ by simply listing the codes of the related elements. Similarly each function can be encoded by a string of length $O(p(|w|)^{l+1})$ and each constant by a string of length $O(p(|w|))$. We can therefore encode an exact description of $\mathfrak{A}$ by a single word $\rho_w \in \{0,1\}^{q(|w|)}$, where $q(n) = O(m \cdot p(n)^l + k \cdot p(n)^{l+1} + r \cdot p(n))$, which is polynomial in the length of $w$.

In a fixed domain $\textbf{dom}(\mathfrak{A})$ a sentence of the form $\forall x \phi(x)$ is true iff the sentence $\bigwedge_{a \in \textbf{dom}(\mathfrak{A})} \phi(a)$ is true. Similarly $\exists x \psi(x)$ is true iff $\bigvee_{b \in \textbf{dom}(\mathfrak{A})} \psi(b)$ is true. Hence to check if: 
$$\forall x_1 \exists x_2 \cdots \forall x_{m-1} \exists x_m \theta(x_1,\hdots,x_m) \in \mathbf{T},$$ 
is true in $\mathfrak{A}$ described by $\rho_w$ it is sufficient to determine whether: 
$$\bigwedge_{a_1 \in \textbf{dom}(\mathfrak{A})} \bigvee_{a_2 \in \textbf{dom}(\mathfrak{A})} \cdots \bigwedge_{a_{m-1} \in \textbf{dom}(\mathfrak{A})} \bigvee_{a_m \in \textbf{dom}(\mathfrak{A})} \theta(a_1,\hdots,a_m),$$ 
is true in $\mathfrak{A}$. This can be achieved by checking whether $\theta(a_1,\hdots,a_m)$ is true in at most $|\textbf{dom}(\mathfrak{A})|^m$ assignments.

There is a fixed number of sentences in $\mathbf{T}$ and the quantifier depth of each one is fixed, hence the time taken to test whether $\mathfrak{A} \models \mathbf{T}$ grows polynomially with $|w|$. As the number of sentences in $\Phi^w_\mathbf{X}$ is equal to $|w|$ and each sentence in $\Phi^w_\mathbf{X}$ is a quantifier-free sentence whose length grows linearly with $|w|$, the time to determine whether $\mathfrak{A}$ models $\Phi^w_\mathbf{X}$ also takes time polynomial in $|w|$.

We can therefore construct a non-deterministic Turing machine $M_1$, that given any input $w \in \Sigma^*$, tries to non-deterministically generate a description $\rho_w$ of some structure $\mathfrak{A}$ modelling $\mathbf{T} \cup \Phi^w_\mathbf{X}$. After generating $\rho_w$ the machine checks in polynomially many steps whether each sentence of $\mathbf{T} \cup \Phi^w_\mathbf{X}$ is true in $\mathfrak{A}$. 
Finally $M_1$ determines whether $\mathfrak{A} \models \Theta$. As $\Theta$ is a fixed finite set of sentences, like $\mathbf{T}$, this decision process can be carried out in time polynomial in $|w|$. If $\mathfrak{A}$ does model $\Theta$ then $M_1$ accepts $w$. If any sentence in $\mathbf{T} \cup \Phi^w_\mathbf{X} \cup \Theta$ is false in $\mathfrak{A}$ then $M_1$ halts. Thus if for all possible $\rho_w$ we have that $\Theta$ is false in any structure which models $\mathbf{T} \cup \Phi^w_\mathbf{X}$ then $M_1$ rejects $w$. By assumption for any $w \in \Sigma^*$, if $\mathfrak{A} \models \mathbf{T} \cup \Phi^w_\mathbf{X}$ then $\mathfrak{A} \models \Theta$ iff $w \in A$. Therefore $M_1$ accepts $w$ if and only if $w \in A$, and as $M_1$ computes in non-deterministic polynomial time we have that $A \in NP$.

Conversely to see that $A \in \text{co-}NP$ we can construct a non-deterministic polynomial time Turing machine $M_2$ which acts the same as $M_1$, except it checks whether $\mathfrak{A}$ models $\Psi$ rather than $\Theta$. By the same reasoning as above $M_2$ accepts $w \in \Sigma^*$ iff $w \in \Sigma^* \setminus A$, therefore $\Sigma^* \setminus A \in NP$ and $A \in \text{co-}NP$. Thus by combining this with the above result we have $A \in NP \cap \text{co-}NP$. 

$(\Leftarrow)$ If $B \in NP \cap \text{co-}NP$ then $B \in NP$ and $\Sigma^* \setminus B \in NP$, hence there must exist two non-deterministic polynomial time Turing machines $N_1,N_2$ that respectively decide $B$ and $\Sigma^* \setminus B$. Without loss of generality, as in Example \ref{TMFFOTME} we can take $N_1$ and $N_2$'s tapes to be infinite in only the rightwards direction. To avoid confusion we can also let $N_1$ and $N_2$ have disjoint sets of internal states. We can then construct a FFOT machine which can implement the rules from either $N_1$ or $N_2$, to decide $B$ as follows.

For $i \in \{1,2\}$ let Turing machine $N_i$ use the alphabet $\Lambda_i \supseteq \Sigma \cup \{\textbf{L},\mathbf{b}\}$, internal states $\Pi_i$ and have initial state $s_{0_i}$ and accepting state $s_{a_i}$. Let $N_i$ have non-deterministic rule set $\mathbf{R}_i$, and for each $(t,b) \in (\Pi_i \times \Lambda_i)$ let $\mathbf{R}_i^{(t,b)}$ denote the set of rules of $\mathbf{R}_i$ prefixed by $(t,b)$. If $N_i$ is in state $t$ reading $b$ then any one of the rules in $\mathbf{R}_i^{(t,b)}$ may be applied. As in Example \ref{TMFFOTME} let $\mathbf{V}^{N_i}$ be the vocabulary used in describing a Turing machine with the above alphabet and state set.

In the vocabulary of $\mathbf{V}^{N_1} \cup \mathbf{V}^{N_2} \cup \{e\}$ let $\mathbf{T}\mathbf{M}'_{N_1,N_2} = (\mathbf{T}\mathbf{M}\mathbf{T}'_{N_1,N_2},\hat{\Sigma}^*_{\mathbf{X}},\{\{I(h) = s_{a_1}\},\{I(h) = s_{a_2}\}\}),$ be a FFOT machine with theory: 
$$\mathbf{T}\mathbf{M}\mathbf{T}'_{N_1,N_2} = \left\{\begin{array}{l} (H(0) = 1) \land (C(0,0) = \textbf{L}), \\
(I(0) = s_{0_1}) \lor (I(0) = s_{0_2}), \\
\forall y (C(0,y) = \mathbf{b}) \rightarrow (C(0,S(y)) = \mathbf{b}), \\
\forall y (\neg(H(x) = y) \rightarrow (C(S(x),y) = C(x,y))), \\
(I(h) = s_{a_1}) \lor (I(h) = s_{a_2})
\end{array}\right\} \cup EQ^=_{\mathbf{V}^{N}} \cup PSA_f \cup \mathbf{\hat{R}}_1 \cup \mathbf{\hat{R}}_2 \cup HT_{(s_{a_1},s_{a_2})}.$$
where for $i \in \{1,2\}$ the non-deterministic rules of $\mathbf{R}_i$ are implemented by:
$$\mathbf{\hat{R}}_i = \bigcup_{(t,b) \in (\Pi_i \times \Lambda_i)}\left\{\forall x \left(\mu_{(t,b)}(x,H(x)) \rightarrow \bigvee_{(t,b;u,c,p) \in \mathbf{R}_i^{(t,b)}}\big(\mu_{(u,c)}(S(x),H(x)) \land \pi_{(p)}(H(x),H(S(x)))\big)\right)\right\}.$$
Where the terms $\mu$ and $\pi$ are as they are in Example \ref{TMFFOTME}. It is then the case that any model $\mathfrak{C}$ of $\mathbf{T}\mathbf{M}\mathbf{T}'_{N_1,N_2} \cup \Phi_X^w$ describes a possible computation path of either $N_1$ or $N_2$. As by the second sentence of $\mathbf{T}\mathbf{M}\mathbf{T}'_{N_1,N_2}$ the model begins in either of the states, and afterwards the sentences of $\mathbf{\hat{R}}_1 \cup \mathbf{\hat{R}}_2$ allow for any one of the appropriate rules to be implemented at each time step of $\mathfrak{C}$. Two different rules cannot be implemented simultaneously as this would lead to a contradiction.

The set $HT_{(s_{a_1},s_{a_2})}$ is as it is in Example \ref{TMFFOTME} with $s_a$ and $s_r$ replaced by $s_{a_1}$ and $s_{a_2}$. Crucially by the fifth sentence of $\mathbf{T}\mathbf{M}\mathbf{T}'_{N_1,N_2}$ any model $\mathfrak{C}$ must reach one of the two accept states. Hence the computation in $\mathfrak{C}$ must be an accepting computation, and if $w \in B$ then $\mathfrak{C}$ must describe a computation of $N_1$ that ends in state $s_{a_1}$, as any computation of $N_2$ on input $w$ would end in the reject state. Conversely if $w \in \Sigma^* \setminus B$ then $\mathfrak{C}$ must describe a computation of $N_2$ that ends in state $s_{a_2}$.

Regardless, this means that $\mathfrak{C} \models (I(h) = s_{a_1})$ iff $w \in B$. We also know that any accepting computation of $N_1$ or $N_2$ takes a polynomial number of time steps. Therefore by our reasoning in Example \ref{TMFMFFOTME} and the fact that $PSA_f \subset \mathbf{T}\mathbf{M}\mathbf{T}'_{N_1,N_2}$ we know that $|\textbf{dom}(\mathfrak{C})|$ may be polynomial in $|w|$. 
Consequently $\mathbf{T}\mathbf{M}'_{N_1,N_2}$ is able to compute $B$ in polynomial resources. \hfill $\Box$
\vspace{2mm}

\noindent
The FFOT machine described in the above proof will only follow a computational path if that path eventually leads to an accept state. The only way the machine could know which paths to take would be if potential future states are somehow able to influence the present states. The machine therefore acts in a non-causal and somewhat atemporal manner, whilst still being clearly bounded in its computational capabilities. 

If $P \not= NP \cap \text{co-}NP$ then our result implies that atemporal/non-causal physical computation is more powerful then classical sequential computation. The problems with known quantum polynomial time algorithms that are believed to lie in $BQP \setminus P$ can all be phrased as a hidden subgroup problem \cite{nielsen2010quantum}, which also lies in $NP \cap \text{co-}NP$.
Therefore our result adds further evidence to the idea that source of the quantum computational speed-up lies in quantum computers being able to act in an atemporal/non-causal manner.

\bibliographystyle{eptcs}
\bibliography{PC2018Bibliography}

\begin{thebibliography}{10}
\providecommand{\bibitemdeclare}[2]{}
\providecommand{\surnamestart}{}
\providecommand{\surnameend}{}
\providecommand{\urlprefix}{Available at }
\providecommand{\url}[1]{\texttt{#1}}
\providecommand{\href}[2]{\texttt{#2}}
\providecommand{\urlalt}[2]{\href{#1}{#2}}
\providecommand{\doi}[1]{doi:\urlalt{http://dx.doi.org/#1}{#1}}
\providecommand{\bibinfo}[2]{#2}

\bibitemdeclare{inproceedings}{baumeler2018computational}
\bibitem{baumeler2018computational}
\bibinfo{author}{{\"A}min \surnamestart Baumeler\surnameend} \&
  \bibinfo{author}{Stefan \surnamestart Wolf\surnameend}
  (\bibinfo{year}{2018}): \emph{\bibinfo{title}{Computational tameness of
  classical non-causal models}}.
\newblock In: {\sl \bibinfo{booktitle}{Proc. R. Soc. A}},
  \bibinfo{volume}{474}, \bibinfo{organization}{The Royal Society}, p.
  \bibinfo{pages}{20170698}, \doi{10.1098/rspa.2017.0698}.

\bibitemdeclare{phdthesis}{Blakey2010model}
\bibitem{Blakey2010model}
\bibinfo{author}{Edward~William \surnamestart Blakey\surnameend}
  (\bibinfo{year}{2010}): \emph{\bibinfo{title}{A model-independent theory of
  computational complexity : from patience to precision and beyond}}.
\newblock Ph.D. thesis, \bibinfo{school}{University of Oxford, {UK}}.
\newblock
  \urlprefix\url{http://ora.ox.ac.uk/objects/uuid:5db40e2c-4a22-470d-9283-3b59b99793dc}.

\bibitemdeclare{article}{blum1989theory}
\bibitem{blum1989theory}
\bibinfo{author}{Lenore \surnamestart Blum\surnameend}, \bibinfo{author}{Mike
  \surnamestart Shub\surnameend}, \bibinfo{author}{Steve \surnamestart
  Smale\surnameend} et~al. (\bibinfo{year}{1989}): \emph{\bibinfo{title}{On a
  theory of computation and complexity over the real numbers: $ NP
  $-completeness, recursive functions and universal machines}}.
\newblock {\sl \bibinfo{journal}{Bulletin (New Series) of the American
  Mathematical Society}} \bibinfo{volume}{21}(\bibinfo{number}{1}), pp.
  \bibinfo{pages}{1--46}, \doi{10.1090/S0273-0979-1989-15750-9}.

\bibitemdeclare{article}{cobham1969intrinsic}
\bibitem{cobham1969intrinsic}
\bibinfo{author}{Alan \surnamestart Cobham\surnameend} \&
  \bibinfo{author}{Yehoshua \surnamestart Bar-Hillel\surnameend}
  (\bibinfo{year}{1969}): \emph{\bibinfo{title}{The intrinsic computational
  difficulty of functions}}.

\bibitemdeclare{book}{cooper2003computability}
\bibitem{cooper2003computability}
\bibinfo{author}{S.~Barry \surnamestart Cooper\surnameend}
  (\bibinfo{year}{2004}): \emph{\bibinfo{title}{Computability theory}}.
\newblock \bibinfo{publisher}{Chapman \& Hall/CRC, Boca Raton, FL}.

\bibitemdeclare{inproceedings}{deutsch1985quantum}
\bibitem{deutsch1985quantum}
\bibinfo{author}{David \surnamestart Deutsch\surnameend}
  (\bibinfo{year}{1985}): \emph{\bibinfo{title}{Quantum theory, the
  Church-Turing principle and the universal quantum computer}}.
\newblock In: {\sl \bibinfo{booktitle}{Proceedings of the Royal Society of
  London A: Mathematical, Physical and Engineering Sciences}},
  \bibinfo{volume}{400}, \bibinfo{organization}{The Royal Society}, pp.
  \bibinfo{pages}{97--117}, \doi{10.1098/rspa.1985.0070}.

\bibitemdeclare{article}{edmonds1965paths}
\bibitem{edmonds1965paths}
\bibinfo{author}{Jack \surnamestart Edmonds\surnameend} (\bibinfo{year}{1965}):
  \emph{\bibinfo{title}{Paths, trees, and flowers}}.
\newblock {\sl \bibinfo{journal}{Canadian Journal of mathematics}}
  \bibinfo{volume}{17}(\bibinfo{number}{3}), pp. \bibinfo{pages}{449--467},
  \doi{10.4153/CJM-1965-045-4}.

\bibitemdeclare{book}{epstein2011classical}
\bibitem{epstein2011classical}
\bibinfo{author}{Richard~L \surnamestart Epstein\surnameend}
  (\bibinfo{year}{2011}): \emph{\bibinfo{title}{Classical mathematical logic:
  the semantic foundations of logic}}.
\newblock \bibinfo{publisher}{Princeton University Press}.

\bibitemdeclare{article}{Gurevich2000}
\bibitem{Gurevich2000}
\bibinfo{author}{Yuri \surnamestart Gurevich\surnameend}
  (\bibinfo{year}{2000}): \emph{\bibinfo{title}{Sequential abstract-state
  machines capture sequential algorithms}}.
\newblock {\sl \bibinfo{journal}{ACM Trans. Comput. Log.}}
  \bibinfo{volume}{1}(\bibinfo{number}{1}), pp. \bibinfo{pages}{77--111},
  \doi{10.1145/343369.343384}.

\bibitemdeclare{article}{hamkins2000infinite}
\bibitem{hamkins2000infinite}
\bibinfo{author}{Joel~David \surnamestart Hamkins\surnameend} \&
  \bibinfo{author}{Andy \surnamestart Lewis\surnameend} (\bibinfo{year}{2000}):
  \emph{\bibinfo{title}{Infinite time Turing machines}}.
\newblock {\sl \bibinfo{journal}{The Journal of Symbolic Logic}}
  \bibinfo{volume}{65}(\bibinfo{number}{2}), pp. \bibinfo{pages}{567--604},
  \doi{10.2307/2586556}.

\bibitemdeclare{inproceedings}{WhenDoesAPhysicalSystemCompute}
\bibitem{WhenDoesAPhysicalSystemCompute}
\bibinfo{author}{Clare \surnamestart Horsman\surnameend},
  \bibinfo{author}{Susan \surnamestart Stepney\surnameend},
  \bibinfo{author}{Rob~C \surnamestart Wagner\surnameend} \&
  \bibinfo{author}{Viv \surnamestart Kendon\surnameend} (\bibinfo{year}{2014}):
  \emph{\bibinfo{title}{When does a physical system compute?}}
\newblock In: {\sl \bibinfo{booktitle}{Proc. R. Soc. A}},
  \bibinfo{volume}{470}, \bibinfo{organization}{The Royal Society}, p.
  \bibinfo{pages}{20140182}, \doi{10.1098/rspa.2014.0182}.

\bibitemdeclare{book}{kaye1991}
\bibitem{kaye1991}
\bibinfo{author}{Richard \surnamestart Kaye\surnameend} (\bibinfo{year}{1991}):
  \emph{\bibinfo{title}{Models of {P}eano arithmetic}}.
\newblock {\sl \bibinfo{series}{Oxford Logic Guides}}~\bibinfo{volume}{15},
  \bibinfo{publisher}{The Clarendon Press, Oxford University Press, New York}.
\newblock \bibinfo{note}{Oxford Science Publications}.

\bibitemdeclare{book}{nielsen2010quantum}
\bibitem{nielsen2010quantum}
\bibinfo{author}{Michael~A. \surnamestart Nielsen\surnameend} \&
  \bibinfo{author}{Isaac~L. \surnamestart Chuang\surnameend}
  (\bibinfo{year}{2000}): \emph{\bibinfo{title}{Quantum computation and quantum
  information}}.
\newblock \bibinfo{publisher}{Cambridge University Press, Cambridge},
  \doi{10.1017/CBO9780511976667}.

\bibitemdeclare{book}{vakil2011real}
\bibitem{vakil2011real}
\bibinfo{author}{Nader \surnamestart Vakil\surnameend} (\bibinfo{year}{2011}):
  \emph{\bibinfo{title}{Real analysis through modern infinitesimals}}.
\newblock \bibinfo{publisher}{Cambridge University Press},
  \doi{10.1017/CBO9780511740305}.

\bibitemdeclare{article}{vergis1986complexity}
\bibitem{vergis1986complexity}
\bibinfo{author}{Anastasios \surnamestart Vergis\surnameend},
  \bibinfo{author}{Kenneth \surnamestart Steiglitz\surnameend} \&
  \bibinfo{author}{Bradley \surnamestart Dickinson\surnameend}
  (\bibinfo{year}{1986}): \emph{\bibinfo{title}{The complexity of analog
  computation}}.
\newblock {\sl \bibinfo{journal}{Mathematics and computers in simulation}}
  \bibinfo{volume}{28}(\bibinfo{number}{2}), pp. \bibinfo{pages}{91--113},
  \doi{10.1016/0378-4754(86)90105-9}.

\bibitemdeclare{inproceedings}{whyman2018physical}
\bibitem{whyman2018physical}
\bibinfo{author}{Richard \surnamestart Whyman\surnameend}
  (\bibinfo{year}{2018}): \emph{\bibinfo{title}{Physical Computation and
  First-Order Logic}}.
\newblock In: {\sl \bibinfo{booktitle}{Machines, Computations, and Universality
  - 8th International Conference, {MCU} 2018, Fontainebleau, France, June
  28-30, 2018}}, \doi{10.1007/978-3-319-92402-1}.

\end{thebibliography}
\renewcommand{\thesection}{\Alph{section}}
\setcounter{section}{0}
\setcounter{ThmNum}{0}
\section{Appendix}
\begin{defn}
\label{EAD}
\normalfont
In a vocabulary $\mathbf{V}$ for each $m$-ary relation $R \in \mathbf{V}$ and $n$-ary function $f \in \mathbf{V}$ let:
\begin{align*}
EQ_R^= &\equiv \forall x_1 \hdots \forall x_m \forall y_1 \hdots \forall y_m \bigwedge_{i=1}^m (x_i = y_i) \rightarrow (R(x_1,\hdots,x_m) \leftrightarrow R(y_1,\hdots,y_m)), \\
EQ_f^= &\equiv \forall x_1 \hdots \forall x_n \forall y_1 \hdots \forall y_n \bigwedge_{i=1}^n (x_i = y_i) \rightarrow (f(x_1,\hdots,x_n) = f(y_1,\hdots,y_n)).
\end{align*}
The \textbf{equality axioms} \cite{epstein2011classical} for the binary relation $=\, \in \mathbf{V}$ are then:
$$EQ^=_\mathbf{V} = \{EQ_V^= \text{ }|\text{ } V \in \mathbf{V}\} \cup \left\{\begin{array}{l}
\forall x (x = x), \\
\forall x \forall y (x = y) \rightarrow (y = x), \\
\forall x \forall y \forall z ((x = y) \land (y = z)) \rightarrow (x = z)
\end{array}\right\}.$$
\end{defn}
\begin{defn}
\label{FOPAD}
\normalfont
Within the vocabulary of $\mathbf{V}^S = \{=,S,0\}$ where $S$ is a unary function, and $0$ is a constant symbol, 
the \textbf{first-order Peano successor axioms} are:
$$PSA = \left\{\begin{array}{l}
\forall x \neg (S(x) = 0), \\ 
\forall x \forall y ((S(x) = S(y)) \rightarrow (x = y)), \\
\forall x \neg (S(x) = x)
\end{array}\right\}.$$
\end{defn}
Any model of $EQ^=_{\mathbf{V}^S} \cup PSA$ is either the usual structure of the natural numbers $\langle \mathbb{N};=,S,0\rangle$ or a structure with an initial segment that is isomorphic to $\langle \mathbb{N};=,S,0\rangle$ \cite{kaye1991}.
\begin{defn}
\label{FORAD}
\normalfont
Within the vocabulary $\mathbf{V}^Q = \{=,<,\leqslant,+,\times,0,1\}$ the \textbf{dense ordered field axioms} are:
$$DOF = \left\{\begin{array}{ll}
\forall x ((x + 0) = x) \land ((x \times 0) = 0), & (0 < 1),  \\
\forall x \forall y \forall z (((x + y) + z) = (x + (y + z))), & \forall x \forall y ((x + y) = (y + x)) , \\
\forall x \forall y \forall z (((x \times y) \times z) = (x \times (y \times z))), & \forall x \forall y ((x \times y) = (y \times x)), \\
\forall x \forall y \forall z (((x + y) \times z) = ((x \times z) + (y \times z))), & \forall x \exists y ((x + y) = 0), \\
\forall x \forall y (0 \leqslant y) \rightarrow (x \leqslant (x + y)), & \forall x (\neg(x = 0) \rightarrow \exists y ((x \times y) = 1)), \\
\forall x \forall y \forall z ((x < y) \land (y < z)) \rightarrow (x < z), & \forall x \neg(x < x), \\
\forall x \forall y \forall z (x \leqslant y) \rightarrow ((x + z) \leqslant (y + z)), & \forall x \forall y (x \leqslant y) \leftrightarrow ((x < y) \lor (x = y)), \\
\forall x \forall y \forall z ((0 < z) \land (x \leqslant y)) \rightarrow ((x \times z) \leqslant (y \times z)), & \forall x \forall y (x < y) \lor (x = y) \lor (y < x)
\end{array}\right\}.$$
\end{defn}
Every model $\mathfrak{R}$ of $EQ^=_{\mathbf{V}^Q} \cup DOF$ contains a subset $G$ which can be embedded into the usual structure of the real numbers $\langle \mathbb{R};=,<,\leqslant,+,\times,0,1\rangle$. There also exists a subset of $G$ which is isomorphic to the usual structure of the rational numbers $\langle \mathbb{Q};=,<,\leqslant,+,\times,0,1\rangle$ \cite{vakil2011real}. Hence $G$ is closed under $+,\times$ and the elements of $\mathbb{Q}$. Therefore, the output of a machine including $DOF$ can be independent of what occurs at any non-standard elements outside of $\mathbb{R}$.
\begin{defn}
\label{FFOPAD}
\normalfont
Within the vocabulary of $\mathbf{V}^S_f = \mathbf{V}^S \cup \{e\}$ where $e$ is a constant symbol,
the \textbf{finite Peano successor axioms} are:
$$PSA_f = \left\{\begin{array}{l}
\forall x \neg (S(x) = 0), \\ 
\forall x \forall y ((S(x) = S(y)) \rightarrow ((x = y) \lor (S(x) = e))), \\
\forall x ((S(x) = x) \leftrightarrow (x = e))
\end{array}\right\}.$$
\end{defn}
$EQ^=_{\mathbf{V}^S_f} \cup PSA_f$ is modelled by any finite structure of the form $\langle \{0,1,\hdots,n\};=,S,0,e\rangle$ where $S^i(0) = i$ and $n = e$. Indeed any model $\mathfrak{N}$ of $EQ^=_{\mathbf{V}^S_f} \cup PSA$ can be converted into a model of $EQ^=_{\mathbf{V}^S_f} \cup PSA_f$ by replacing the subset of $\mathfrak{N}$ that is isomorphic to $\langle \mathbb{N};=,S,0\rangle$ with $\langle \{0,1,\hdots,n\};=,S,0,e\rangle$. Alternatively, by disjointly combining the domain of $\mathfrak{N}$ with the set $\{\hdots,-2,-1,e\}$ where $S^{-i}(i) = e$, we also obtain a model of $EQ^=_{\mathbf{V}^S_f} \cup PSA_f$, but with an initial segment that is isomorphic to $\langle \mathbb{N};=,S,0\rangle$.

In a similar manner we can finitely approximate $\mathbb{R}$ via a structure with greatest and least elements given by $e$ and -$e$, as well as error and precision given by $\frac{1}{r}$ and $r$ respectively. To enable an approximate version of multiplication we introduce the approximate equality relation ``$\approx$", which holds if two numbers are within a distance of $\frac{1}{r}$ from one another.
\begin{defn}
\label{FFORAD}
\normalfont
Within the vocabulary $\mathbf{V}^Q_f \cup \{\approx,e,\text{-}e,r,\frac{1}{r}\}$ where $\approx$ is a binary relation and $e,\text{-}e,r,\frac{1}{r}$ are constants let:
$$DOF' = \left\{\begin{array}{l}
(0 \leqslant \frac{1}{r}) \land ((r \times \frac{1}{r}) = 1) \land ((r \times r) = e) \land ((\text{-}e + e) = 0) \land ((\text{-}r + r) = 0), \\
\forall x \forall y (x \approx y) \leftrightarrow ((x \leqslant (y + \frac{1}{r})) \land (y \leqslant (x + \frac{1}{r}))), \\
\forall x (0 \leqslant x) \rightarrow ((e + x) = e), \\
\forall x (1 \leqslant x) \rightarrow ((e \times x) = e), \\
\forall x ((x \leqslant e) \land (\text{-}e \leqslant x)), \\
\forall x \forall y \forall z ((\text{-}e < (x + y),(y + z) < e) \rightarrow (((x + y) + z) = (x + (y + z))), \\
\forall x \forall y \forall z (\text{-}e < (x \times y),(y \times z) < e) \rightarrow (((x \times y) \times z) \approx (x \times (y \times z))), \\
\forall x \forall y \forall z (\text{-}e < (x + y),(x \times z),(y \times z) < e) \rightarrow (((x + y) \times z) \approx ((x \times z) + (y \times z)))
\end{array}\right\}.$$
Where $a \leqslant b,c \leqslant d$ is shorthand for $(a \leqslant b) \land (b \leqslant d) \land (a \leqslant c) \land (c \leqslant d)$.
The \textbf{finite dense ordered field axioms} are then:
$$DOF_f = DOF' \cup DOF \setminus \left\{\begin{array}{l}
\forall x \forall y \forall z (((x + y) + z) = (x + (y + z))), \\
\forall x \forall y \forall z (((x \times y) \times z) = (x \times (y \times z))), \\
\forall x \forall y \forall z (((x + y) \times z) = ((x \times z) + (y \times z)))
\end{array}\right\}.$$
\end{defn}
$EQ^=_{\mathbf{V}^S_f} \cup DOF_f$ is modelled by any finite structure of the form $\langle\{\frac{a}{m} \text{ }|\text{ } a \in \{-m^3,\hdots,m^3\}\};=,<,\leqslant,S,+,\times,$ $0,1,e,\text{-}e,r,\frac{1}{r}\rangle$ where $<,\leqslant,0,1$ are as they usually are in $\mathbb{Q}$. Also $e = \frac{m^3}{m}$, $\text{-}e = \frac{-m^3}{m}$, $r = \frac{m^2}{m}$, and $\frac{1}{r} = \frac{1}{m}$. Addition is as usual with $\frac{a}{m} + \frac{b}{m} = \frac{m^3}{m}$ if $a + b \geqslant m^3$ and $\frac{a}{m} + \frac{b}{m} = -e$ if $a + b \leqslant m^3$. Whereas multiplication is such that $\frac{a}{m} \times \frac{b}{m}$ is approximately equal to whichever element in the domain is nearest to $\frac{ab}{m^2}$.

$EQ^=_{\mathbf{V}^S_f} \cup DOF_f$ may also be modelled by structures with subsets that are isomorphic to the reals and the rationals. This is due to the fact that it is possible that $r$ could be transfinite and $\frac{1}{r}$ could be an infinitesimal, in which case addition and multiplication should act as usual. 
\end{document}